# TOWARDS UNDERSTANDING THE NUCLEUS OF M31

Philippe Crane

European Southern Observatory, Karl-Schwarzschild Strasse 2, D85748 Garching,

Germany

Electronic mail: crane@eso.org

Running Title: M31 NUCLEUS

# ABSTRACT

A simple model of the nucleus of M31 based on *HST* images and ground based spectroscopy is used to investigate the properties of the double nucleus in M31. The model reproduces the general properties observed in the nucleus of M31. In particular, it produces symmetric rotation curves, and shifts of the velocity dispersion peak in the sense observed. Within the constraints of the model, the brightest point in the *V* band in the nucleus is confirmed to be physically near to the dynamical center of M31 and not seen in projection along the line of sight. The velocity and velocity dispersion of this object can be constrained. This implies that P2 is colliding with M31 and likely to be in the process of being tidally disrupted.

*Subject headings:* galaxies: individual: M31 galaxies: nuclei galaxies: stellar content

## 1. INTRODUCTION

The observations of the nucleus of M31 present several anomalies which have only recently begun to be recognized for the true puzzles which they contain. The double nature of the nucleus was first definitively revealed by the HST observations of Lauer *et al.*(1993). Subsequently, King, Stanford, and Crane(1995) showed, also from HST data, that the brightest nuclear point in the V and R images of Lauer *et al.*(referred to as P1 in the following) was not the brightest point in the ultraviolet at 1750Å but rather the weaker source in V (referred to as P2) was the brightest. The UV bright P2 lies close to the center of the outer isophotes, and is presumably at the dynamical center of the galaxy. P1 has been interpreted as a body which is merging with the M31 nucleus (Lauer *et al.*1993). Some support for this hypothesis can be found in the fact the P1 has the same color as the reddened "bowl" of stars surrounding the nucleus. Indeed, P1 is not seen in the color difference image of King, Stanford, and Crane (1995) and thus matches very closely the color of the "bowl". Neither P2 nor the "bowl" match the color of P1 or the region outside about 1.8 arcsec where the bulge proper may be thought to begin. Further support that P1 is colliding with M31 is found in the decomposition of King, Stanford, and Crane which does not require any dust in order to reproduce the observed light profiles. On the other hand, the dynamical arguments given by Lauer *et al.*(1993) imply that if P1 is merging with P2, then we are seeing it a very special time.

However, if the imaging data are difficult to understand, the spectroscopy is even more difficult. The best spectroscopic study in the literature to date is the work of Bacon *et al.*(1994). These data show that the rotation curve is centered on the UV bright peak(P2) found by King, Stanford, and Crane(1995), and shows no evidence of P1 inspite of the fact the the data were obtained at wavelengths where P1 is the brightest point in the galaxy. Furthermore, the peak in the velocity dispersion lies neither at P1, or P2, nor between them, but along the line joining them near to P2, but on the side away from P1. This is a most puzzling state of affairs.

It is the purpose of this note to suggest an explanation of the observations and, in particular, an explanation for the anomaly seen in the velocity dispersion. Given the new understanding of the light distribution from the HST observations(Lauer *et al.*(1993), Crane *et al.*(1993) and King, Stanford, and Crane(1995)), the model used in the following discussion is now on somewhat more secure grounds than previous models( Kormendy 1988; Dressler and Richstone 1988).

Previously, Bacon *et al.*(1994) have proposed a similar explanation to the one here, but did not explore the parameter space very extensively. Lauer *et al.*(1993) mention that the effects of a low velocity dispersion component such as P1 will always be visible in the data. More recently, Tremaine(1995) has shown that an eccentric disk model of the M31 nucleus can reproduce some of the observed anomalies.

The next section presents a simple descriptive model of the imaging and spectroscopic data. Observations of the model with different parameters are presented in section 3, section 4 presents the results, and the final section discusses the implications of these results.

## 2. A DESCRIPTIVE MODEL OF THE NUCLEUS OF M31

The results from King, Stanford, and Crane(1995), from Lauer *et al.*(1993), and from Bacon *et al.*(1994) are used to create a simplified description of the nuclear region of M31. This description is intended to reproduce the general properties seen in the observations of the M31 nucleus, but not to reproduce the details. The model consist of two parts: a light distribution, and a model for the spectrum. The procedure described below is similar to that followed by Kormendy(1988) except that in this case the spectra are postulated and not synthesized from stellar templates.

The light distribution in the nuclear region of M31 is taken to be the sum of 2 symmetric Hubble profiles with the same central intensity, and different values for the width parameter. The equation below was used.

$$I(r; I_0, b_1, b_2, a) = \frac{I_0}{(1.0 + \frac{r}{b_2})^2} + \frac{I_0}{(1.0 + \frac{(r-a)}{b_1})^2}$$

where $I_0$ is the central intensity, $b_1$ is the width parameter for P1 and $b_2$ is the width parameter for P2, and "a" is the separation of P1 from P2. The value of $I_0$ is rather arbitrary but is the same for both P1 and P2 as found by King $et$ $al.$(1995). The values which set the linear scales were chosen so that $b_2$ was 2 arc-sec, $b_1$ was chosen as 0.4 arc-sec, and the separation parameter "a" was set to 0.5 arc-sec. These values match reasonably closely the values determined by Lauer $et$ $al.$(1993) and by King, Stanford, and Crane (1995). The profiles and the isophotal contours look similar to the HST data.

The model for the spectra was a bit more complicated in that it had to mimic both the velocity dispersion and the rotation curve observations. The velocity dispersion for P2 followed very closely the suggestion of Bacon $et$ $al.$(1994). Thus the P2 velocity dispersion was chosen to be a constant of 160 $kms^{-1}$ with a Gaussian shaped increase in the central region at P2. The magnitude and width of this increase were variable parameters, and were chosen close to the values suggested by Bacon $et$ $al.$(1994). Thus the height was 20 $km$ $s^{-1}$ and the full width was 0.25 arcsec.

For P1, the velocity dispersion, $\sigma_{P1}$, was varied in the range from 20 to 120 km $s^{-1}$ and the velocity relative to the center of P2, $V_{P1}$, was also varied. P1 does not rotate in this model.

The model for the rotation of the P2 component was chosen to have the same general features as the rotation map and rotation curve given by Bacon *et al.* (1994) (See their Fig. 14 and 16 ). The model for the nuclear region rotates around the central peak, P2, with a rotation axis perpendicular to the line joining P1. The rotation was entirely symmetric around P2. The peak amplitude of the rotation ($\approx 200$ km s$^{-1}$) and the distance between the peaks ($\approx 1.4$ arc-sec) were varied within a range that gave results compatible with the observed rotation curve given by Bacon *et al.*(1994). The rotation declined linearly above and below the line joining P1 and P2 and had a Keplerian fall off outside the peaks in agreement with the observed rotation map. Within the limitations of the model, the peak in the rotation curve, and the distance between the peaks (the steepness) are constrained to lie close to the values given above in order to agree with the observations and to produce the observed effects.

To complete the model, each pixel for each component of the light profile was assigned a spectrum which was determined according to the procedure set out above. The spectrum at any given pixel thus had a continuum brightness determined by the light profile at that point, and a Gaussian absorption profile whose width and position were determined by the above prescription. The depth of the absorption feature for P1 was 0.6 compared to 0.4 for P2. Thus a 3-dimensional array was generated with two spatial dimensions, and one wavelength dimension. At each wavelength point, the spatial dimensions were convolved with a Gaussian seeing function of full width at half maximum 0.8 arc-sec. The spectra were sampled every 10 km s$^{-1}$, and the spatial sampling was 0.08 arc-sec.

## 3.  OBSERVING THE MODEL

The next step was to observe the models. This was done by fitting a single Gaussian absorption line to the spectra along the line joining P1 with P2. For each position, the center of the Gaussian and the width were determined. The centers were plotted to provide a rotation curve, and the widths were plotted to give the velocity dispersion as a function of position. This procedure is a reasonable approximation to the process of finding the broadening curve. It works well here because there is no noise in the simulated spectra.

Fig. 1 shows the rotation curve and the velocity dispersion plot before and after convolution with a seeing function for a model which included P2 only. The main effect of seeing is on the velocity dispersion where there is a large amount of rotational mixing evident.

The addition of P1 to the model increases the range of phenomena that can be produced. Fig. 2 shows the same model as in Fig. 1 with the addition P1. In the model of Fig. 2, $V_{P1} = 80$ km $s^{-1}$ and $\sigma_{P1} = 80$ km $s^{-1}$. These are values which are representative of the values which produce velocity dispersion curves that agree with the observed data. Since P1 is at most only about twice as bright as the stars associated with P2 at the position of P1, P1 has only a very minor effect on the rotation curve. In fact, for a wide range of values of $V_{P1}$, and $\sigma_{P1}$, the models produce symmetric rotation curves with no obvious effect of P1.

For the velocity dispersion curves, the situation is considerably different and more complicated. The effect of P1 is more pronounced as seen in solid line of Fig. 2b. The combination of the low velocity dispersion of P1 and the higher velocity dispersion of P2 combined with the velocity of P1 relative to P2 produce a variety of effects on the un-convolved velocity dispersion shown by the dashed line in Fig 2b. In this particular case, the velocity of P1 relative to P2 has the effect of broadening the observed line and of narrowing it in the region on either side of P1. Shifting $V_{P1}$ from 80 km s$^{-1}$ to 130 km s$^{-1}$ eliminates the peak and accentuates the dip in the un-convolved velocity dispersion curve. The solid line in Fig. 2b shows the effects of convolution with a seeing function. This latter curve is the one of main interest here, because it is this which is to be compared to the observations.

Fig. 3 shows the effect on the velocity dispersion curve of varying the the value of $V_{P1}$ from 0 km s$^{-1}$ to 200 km s$^{-1}$. Clearly neither of these curves would match very well the data as show in the following discussion.

Fig. 4 shows a comparison of the model with the data of Bacon *et al.*(1994). The rotation curve shown in the upper panel is quite a good match to the data. The velocity dispersion curve is similar to the velocity dispersion curve of Bacon *et al.*, but the shift of the peak to the right away from P2 is smaller than seen in the observations of Bacon *et al.*(1994). The model shows a dip in the velocity dispersion at about −1.5 arc-sec as seen in the data, and it shows a generally lower dispersion to the left of P2 compared the region to the right of P2. This effect may also be present in the observations.

## 4. RESULTS

The main result of the above analysis is to show that this model can reproduce the general features seen in the spectral observations: the symmetry or near symmetry of the rotation curve, and the apparently anomalous shift of the peak in the velocity dispersion curve. A wide range of possibilities for $V_{P1}$ and for $\sigma_{P1}$ give rotation curves which are compatible with the data. The range of possibilities that give velocity dispersion curves that match the data is more restricted. $\sigma_{P1}$ values in the range between 25 km s$^{-1}$ and 100 km s$^{-1}$ are acceptable, and the middle of this range is preferred. The range for $V_{P1}$ is more difficult to constrain. Setting $V_{P1}$ equal to the co-rotation velocity of 140 km s$^{-1}$ produces too small a shift in the peak of the velocity dispersion curve to match the data. Setting $V_{P1}$ to the velocity of P2, produces a double peaked velocity dispersion curve for $\sigma_{P1} = 80$ km s$^{-1}$ (see Fig. 3). Therefore for the range of parameter space explored, $V_{P1}$ values around $70 \pm 20$ km s$^{-1}$ are needed.

The model has some clear limitations in reproducing the observations. In particular, the shift in the peak of the velocity dispersion curve can only be as large as the half width of the rotational broadening curve. This latter has been chosen to be 0.7 arc-sec in order to match the observed rotation curve, and thus the shifts of the peak in the velocity dispersion observed in these models are generally smaller than this. Since the shift observed by Bacon *et al.*(1994) is close to 0.5 arc-sec, and since the model does not seem to be able to produce shifts this big, this appears to be a limitation of these models in describing this particular data set. However, Dressler and Richstone (1988)( see their Fig. 7) show a somewhat smaller shift than Bacon *et al.*(1994). Preliminary results of Kormendy and Bender(1995) also show a smaller shift in the peak of the velocity dispersion than seen by Bacon *et al.*(1994). In fact the model can quite easily reproduce the Kormendy and Bender(1995), so far unpublished, results. In addition, the range of parameter space that has been explored for the current rather primitive model is limited, and, importantly, the size of the shift depends on the details of the model. Finally, as already mentioned, the observations are not yet of sufficient precision or adequately sampled to expect any model to match precisely.

## 5. DISCUSSION

The analysis here together with the analysis of King, Stanford, and Crane(1995) provides convincing evidence that P1 is indeed dynamically associated with P2 and not merely a superposition. The colors of P1 match those of the UV dip surrounding P2, and indicate that the stars in P1 and in the UV dip come from same population. The constraints on $V_{P1}$ and $\sigma_{P1}$ required to match the spectral data also indicates that P1 is close its projected distance from P2.

From what is now known, it is safe to say that many of the previously postulated hypotheses concerning the nucleus of M31 are unlikely to be correct. Indeed, all models invoking a non uniform dust distribution are not viable (Nieto *et al.*,1986; Dressler and Richstone, 1988; Lauer *et al.*,1993). Kormendy(1988) proposed a disk model. Tremaine(1995) has taken up this proposal recently and this idea is discussed below. Dressler and Richstone(1988) advanced 4 hypothesis. The proposal that P2 is the true center and that P1 is a low M/L substructure orbiting P2 may be supported by the model presented here. Dressler and Richstone's proposal that the M31 nucleus can be modelled by an asymmetric population of stars is similar to the suggestion of Tremaine(1995) discussed in more detail below.

A recent discussion of the anomalies in the nucleus of M31 by Tremaine(1995) proposes that an eccentric disk of stars orbiting P2 can explain the light profile and the spectral data. This model is the latest and most convincing of the disk models for the M31 nucleus and presumably supersedes the previous suggestions of disk models for the M31 nucleus some of which were mentioned above. This disk model is particularly appealing because it avoids one of the central difficulties found in the current proposal; that P1 is in the process of being tidally disrupted and that the probability of seeing this event just now is small. In comparing Tremaine's model to the current work, the isophotal contours of the eccentric disk model have a considerably higher ellipticity than seen in the data. This model also does not account for the symmetry of the UV dip surrounding P2. It also does not do quite as well in matching the spectral data as the current model. Nevertheless, it remains a viable alternative.

The implications of the present model, that we are seeing P1 in the process of being tidally disrupted, have been discussed by several authors and, in particular, by Lauer *et al.*(1993). The limits set here on $V_{P1}$ and on $\sigma_{P1}$ strengthen the conclusions from the previous remarks concerning dynamical friction, the life time of P1 and the $M/L$ ratio of P1. However, the constraint on $\sigma_{P1}$ favors high, but not exceptionally high, mass-to-light ratios for P1; somewhere in the range for $M/L \approx 15 - 40$. This may imply that both P1 and P2 have very deep gravitational potentials.

It is important to re-emphasize that the model presented here is a descriptive model of the data and is not a self-consistent dynamical model. In order to make a self-consistent dynamical model, it would be necessary to tie the mass density implied by the light distribution to the velocity dispersion and rotation curves. Such models have been discussed by Satoh(1980), and by Binney, Davies, and Illingworth(1990) so there is little doubt that such models can easily be constructed. However, a self-consistent dynamical model is not required to explain the features of the observations that this paper addresses.

Unfortunately there can be no conclusions at this time. If the speculations based on this model are correct, the implied observational task is formidable. In order to verify the details it is necessary to measure the velocity and the velocity dispersion of P1. Although P1 is relatively bright having a V magnitude of around 14.5 (King, Stanford, and Crane, 1995), it is very compact and sits on top of an almost equally bright background with a different velocity dispersion and a different mean velocity. However, the difference in the mean velocity is not enough to resolve the feature of P1 from P2. Thus observations will require a spatial resolution of around 0.1 arc-sec and velocity resolution of about 25 km $s^{-1}$ as well as high signal-to-noise. This is not a job for the weak at heart.

It is a pleasure to acknowledge discussions with or comments from R. Bender, I.R. King, G. Monnet and S. Tremaine.

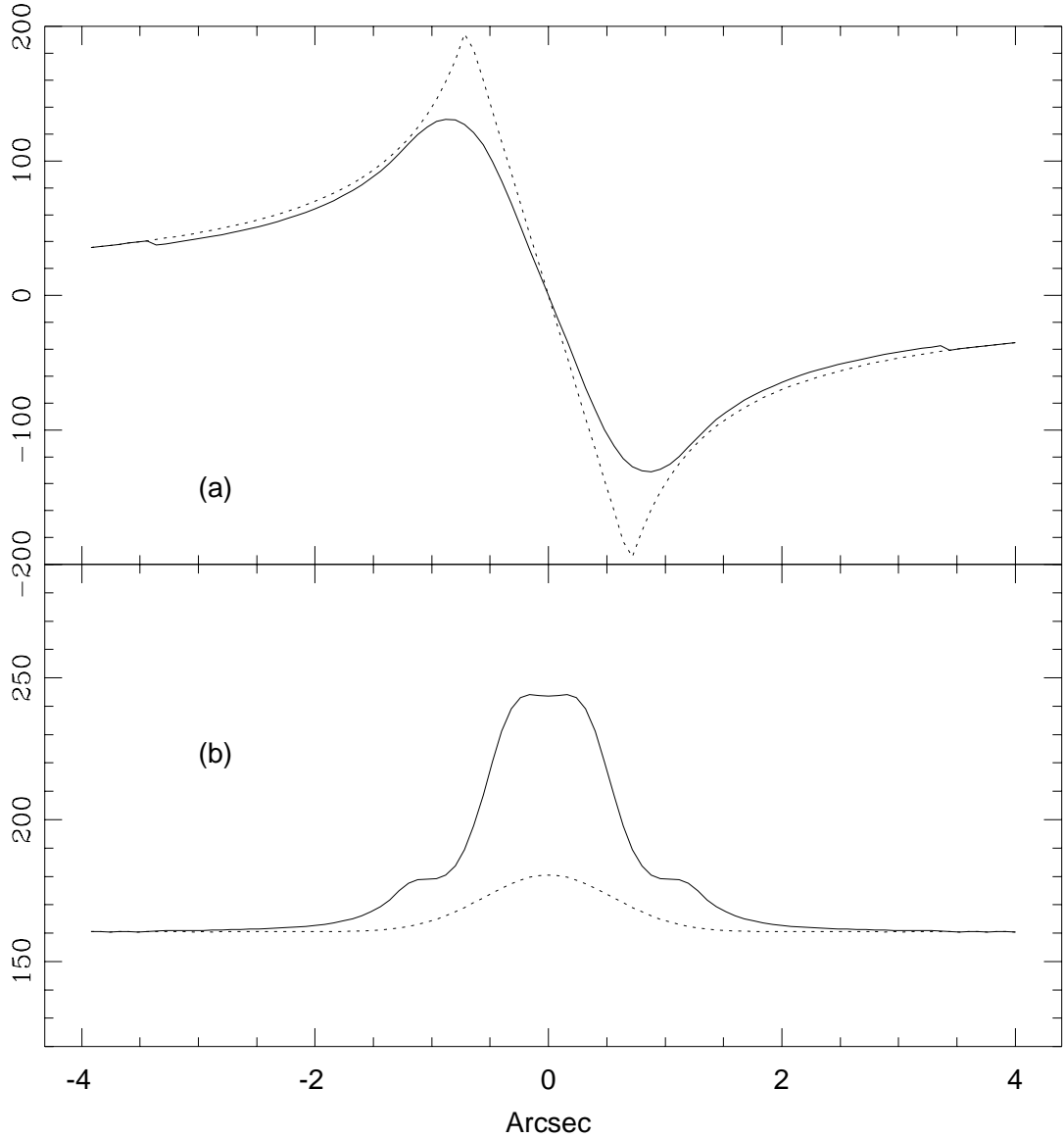

Fig. 1.— Top panel shows the rotation curve which has been extracted from the model along the P1-P2 axis. P2 is centered at 0.0 arc-sec. The units are km s$^{-1}$ The dotted lines are show the rotation curve before convolution with a seeing function, and the solid lines show the rotation after convolution with the seeing. The bottom panel shows the velocity dispersion curve that has been extracted from the same model as for the top panel. The data are taken again along the axis joining P1 to P2. The dotted(solid) lines are before(after) convolution with the seeing function. The effects of rotational mixing are clearly seen.

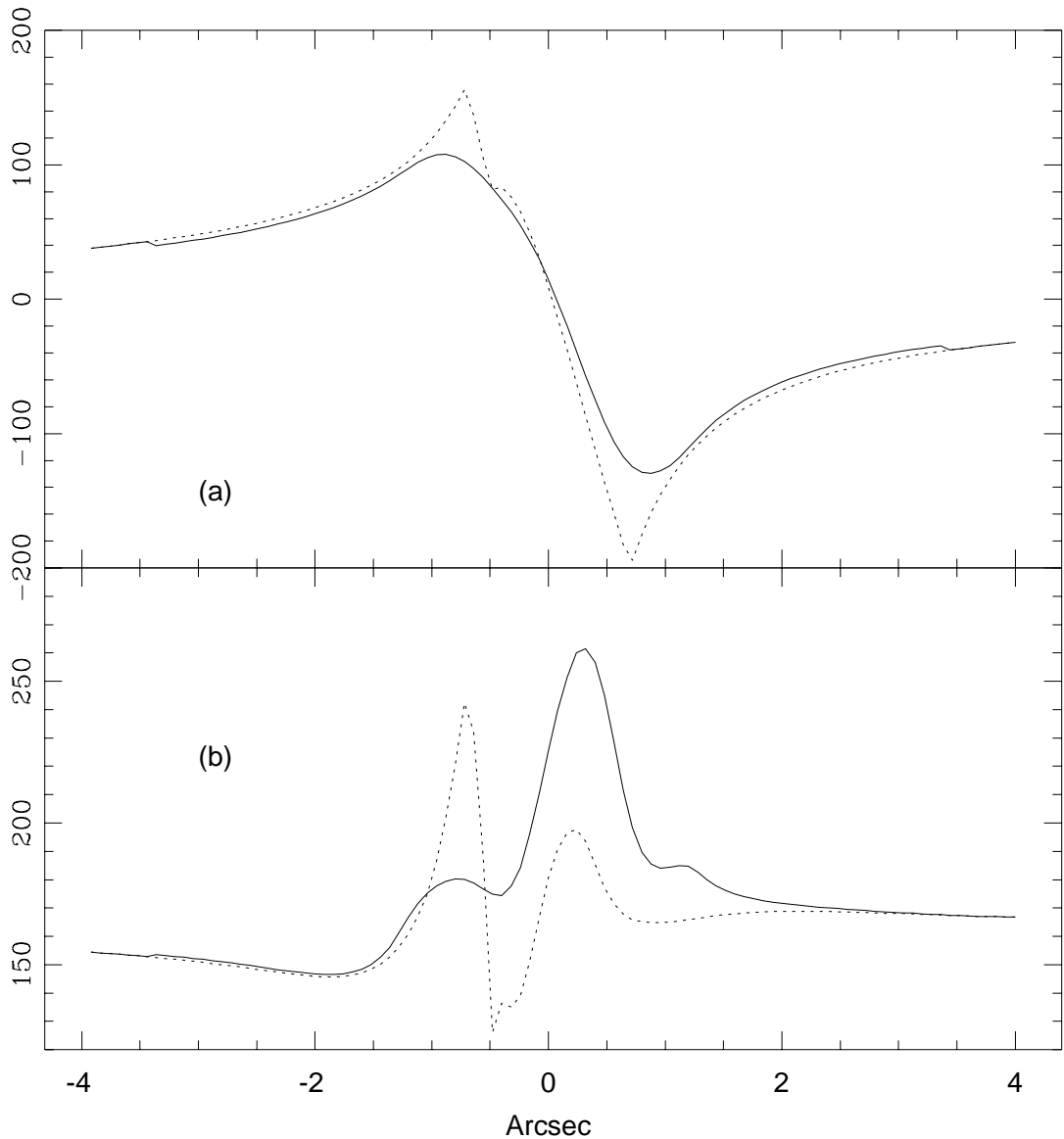

Fig. 2.— The same model as in Fig. 1 but now including P1 as described in the text. P1 is at −0.5 arc-sec in this figure.

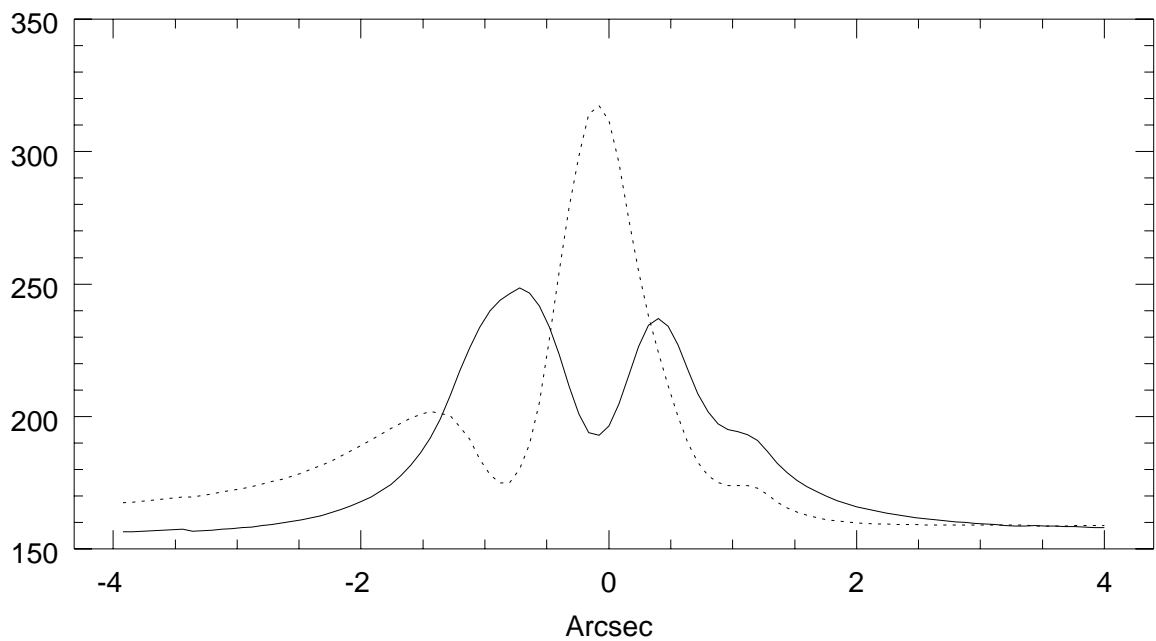

Fig. 3.— Velocity dispersion curves for the same model as Fig. 2, but using different values for $V_{P1}$. The solid curve is for $V_{P1} = 200$ km s$^{-1}$. The dashed curve is for $V_{P1} = 0$ km s$^{-1}$.

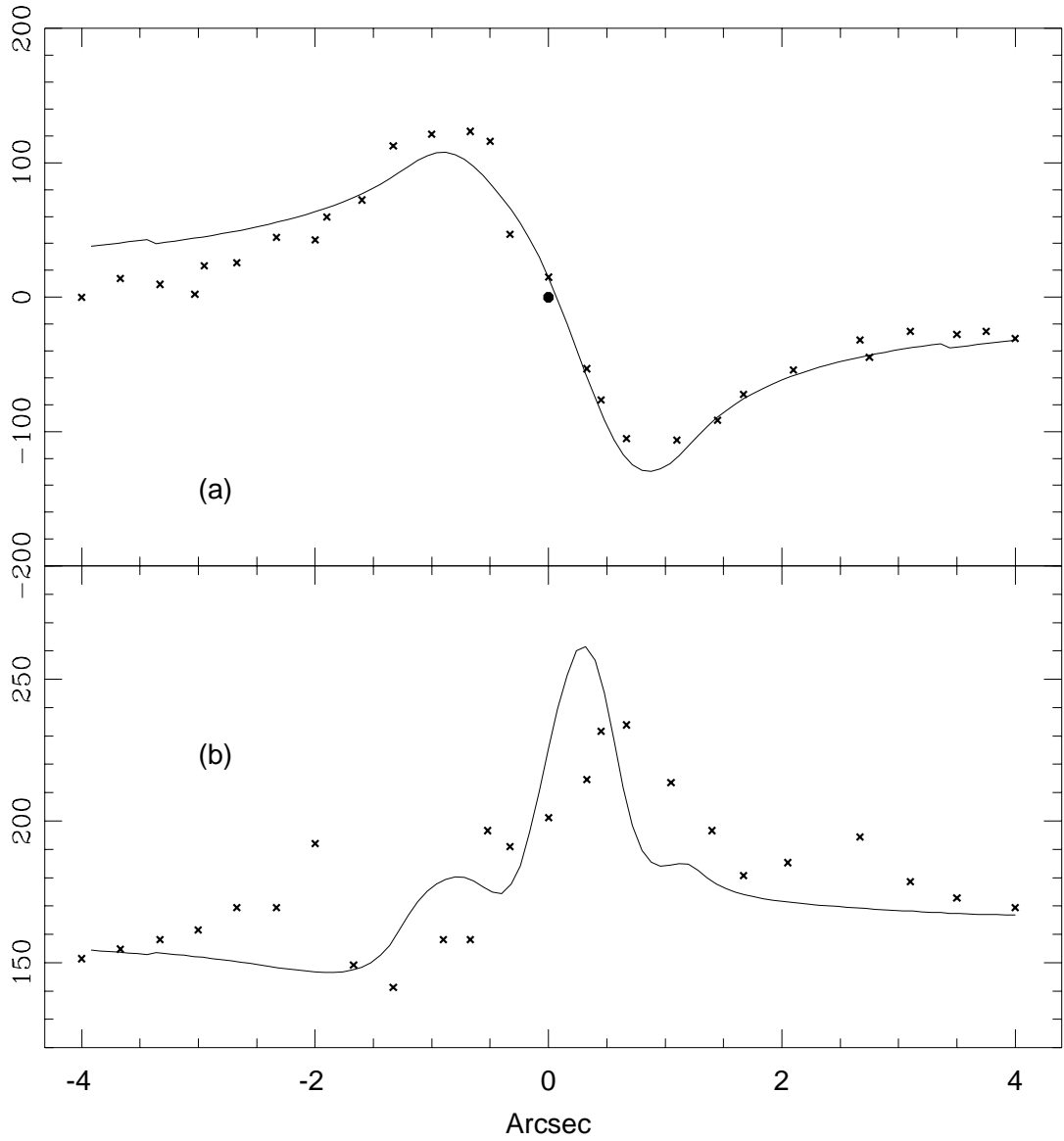

Fig. 4.— Comparison of the model in Fig. 2 with the data data of Bacon *et al.*(1993). The upper panel shows the rotation curve, and the lower panel shows the velocity dispersion. The crosses are taken from the data papers of Bacon *et al.*. The circular dot in the upper panel marks the center of the plot.